# A Novel Hexpyramid Pupil Slicer for an ExAO Parallel DM for the Giant Magellan Telescope

Maggie Kautz[1,2], Laird M. Close[2], Alex Hedglen[1,2], Sebastiaan Haffert[2], Jared R. Males[2], Fernando Coronado[2]

[1]James C. Wyant College of Optical Sciences, University of Arizona, 1630 E University Blvd, Tucson, AZ 85719.

[2]Steward Observatory, University of Arizona, 933 N Cherry Ave, Tucson, AZ 85719.

## ABSTRACT

The 25.4m Giant Magellan Telescope (GMT) will be amongst the first in a new series of segmented extremely large telescopes (ELTs). The 25.4 m pupil is segmented into seven 8.4 m circular segments in a flower petal pattern. At the University of Arizona we have developed a novel pupil slicer that will be used for ELT extreme adaptive optics (ExAO) on the up and coming ExAO instrument, GMagAO-X. This comes in the form of a six-sided reflective pyramid with a hole through the center known as a "hexpyramid". By passing the GMT pupil onto this reflective optic, the six outer petals will be sent outward in six different directions while the central segment passes through the center. Each segment will travel to its own polarization independent flat fold mirror mounted on a piezoelectric piston/tip/tilt controller then onto its own commercial 3,000 actuator deformable mirror (DM) that will be employed for extreme wavefront control. This scheme of seven DMs working in parallel to produce a 21,000 actuator DM is a new ExAO architecture that we named a "parallel DM," in which the hexpyramid is a key optical component. This significantly surpasses any current or near future actuator count for any monolithic DM architecture. The optical system is designed for high-quality wavefront ($\lambda/10$ surface PV) with no polarization errors and no vignetting. The design and fabrication of the invar mechanical mounting structure for this complex optical system is described in this paper.

## 1. INTRODUCTION

The Giant Magellan Telescope will be the first in a new line of extremely large telescopes (ELTs). With seven 8.4 m segments arranged in a flower petal pattern, and an adaptive secondary held up by struts there will be plenty of discontinuities in the pupil, as clear from Figure 1.

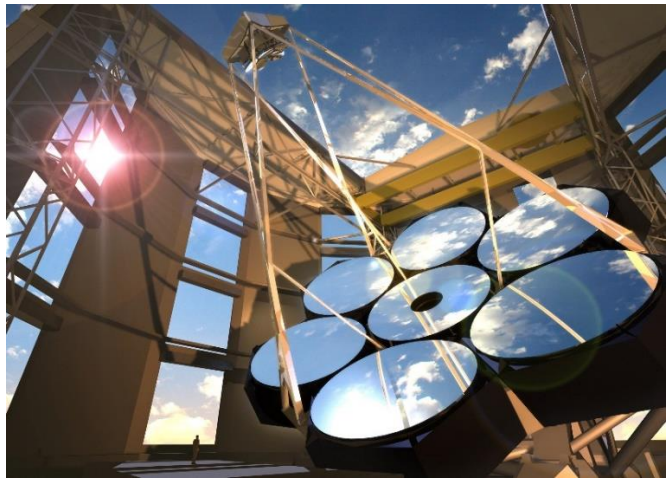

Figure 1. Courtesy: Giant Magellan Telescope – GMTO Corporation

According to the U.S. Decadal Survey on Astronomy and Astrophysics 2020, the largest technical challenge facing the GMT will be co-phasing all seven segments to get a high-Strehl (>70%) point spread function (PSF). As a part of a risk reduction program for the Giant Magellan Telescope, the University of Arizona Center for Astronomical Adaptive Optics (CAAO) and the Extreme Wavefront Control Lab (XWCL) has developed the High Contrast Adaptive-optics Testbed (HCAT, PI Laird Close), which will practice co-phasing the seven mirror segments of the GMT via piston control (Hedglen, et al. 2022)[1,2]. A current extreme adaptive optics (ExAO) system, known as the Magellan Extreme Adaptive Optics system (MagAO-X, PI Jared Males), run by the XWCL was designed for and operates at the 6.5-meter Clay Magellan Telescope at the Las Campanas Observatory in Chile[3,4]. It employs a pyramid wavefront sensor and a 2,040-actuator deformable mirror (DM) operating at 3.63kHz. This has been very successful, producing high-resolution, high-Strehl images[5], but this system is working to correct a 6.5-meter pupil.

The GMT will consist of seven 8.4 m diameter primary segments and have a total diameter of 25.4 meters. An individual DM for each segment of the GMT is required to create the high-quality science images created with MagAO-X. HCAT will investigate the idea of implementing multiple deformable mirrors (DMs) working in parallel, one for each mirror segment of the GMT, to keep the segments in-phase and with low wavefront error. This will involve splitting up the pupil into each of its individual segments and assigning a piezoelectric PI S-325 actuator and deformable mirror to each segment. Prior to purchasing real DMs, the parallel DM architecture will be modeled with "mock DMs" to test whether it is possible to split up the GMT pupil into its seven segments then coherently recombine the beams and control piston to the tens of nanometers level via piezoelectric controllers. This new ExAO architecture requires a novel pupil slicer, described in this paper. The CAAO group hopes to employ the pupil slicer and parallel DM architecture on the up-an-coming visible to near-IR ExAO instrument, GMagAO-X[6]. For more details about GMagAO-X's opto-mechanics please see Close, et al. 2022[6], for the overall science and electronics see Males, et al. 2022[7], and for the wavefront sensing architecture see Haffert, et al. 2022[8].

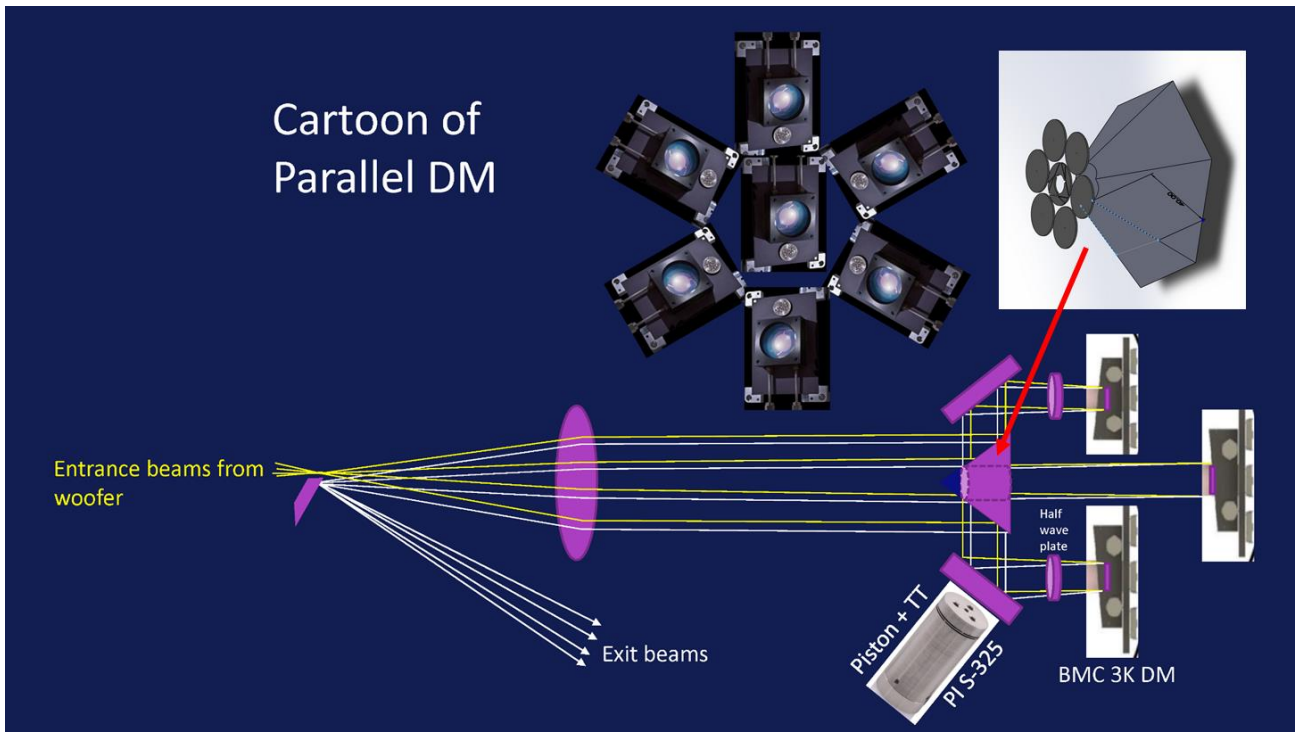

Figure 2. Parallel DM Concept: GMT pupil incident on hexpyramid with each segment being sent to its own PI S-325 controller and BMC 3k DM.

## 2. HEXPYRAMID AND SPLITTING UP THE PUPIL

### 2.1 Optical Design and Test

This pupil slicer comes in the form of a six-sided reflective pyramid with a hole through the center known as a "hexpyramid" (Fig. 3). The hexpyramid has a centerline-side angle of 45 degrees so it can reflect light outward at a 45-degree angle (Fig. 4). Rocky Mountain Instrument Co. was successfully able to manufacture this optic (Fig. 5). By passing the GMT pupil onto this reflective optic, the six outer petals will be sent outward in six different directions while the central segment passes through the center. This optical design included very tight tolerances, laid out in Table 1, due to the wavefront error budget of the testbed.

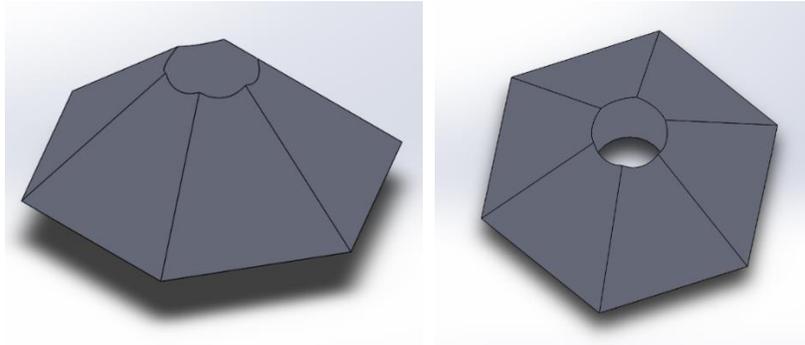

Figure 3. SolidWorks rendition of hexpyramid

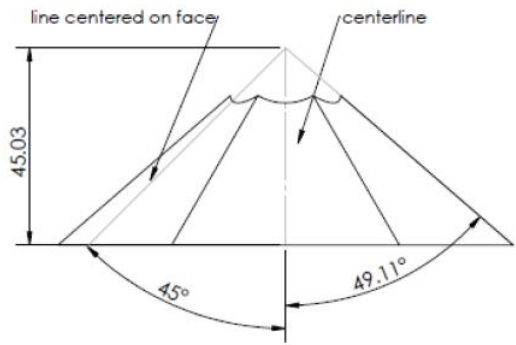

Figure 4. Hexpyramid Drawing

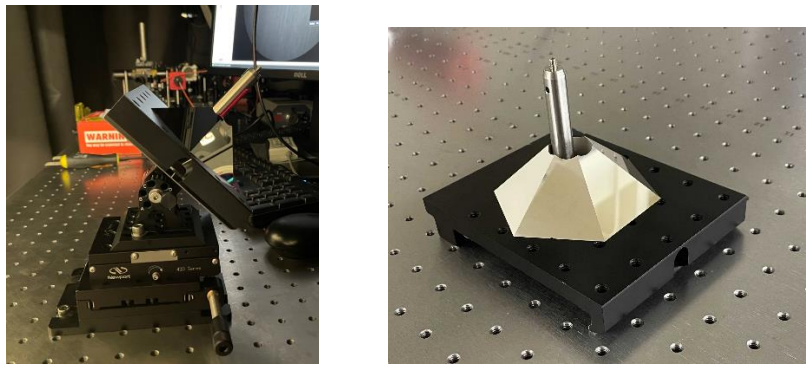

Figure 5. Hexpyramid in-lab: (left) being measured on 45-degree mount in front of a Zygo ®, (right) testing mounting possibilities

Table 1. Hexpyramid Specifications Summary

| Optical Parameter | Specification |
|---|---|
| Material | Fused Silica |
| Waveband | 0.45 – 2.00 μm |
| Coating | Protective Silver, AOI = 45°, $R_{avg} \geq 97\%$ over required waveband |
| PV Surface Flatness in CA (λ at 632.8 nm) | < or = λ/10<br>Achieved ~λ/16 (measured via Zygo ® in Fig. 7) |
| Surface Quality (scratch-dig) in CA | < 20-15 |
| RMS Surface Roughness in CA | < 4 nm |
| Chamfer on clear aperture edge in CA & around hole edges | < 50 μm |

The specifications laid out in the above table apply mainly to the clear aperture on each side of the hexpyramid. The actual beam footprint from the GMT pupil is shown in Figure 6a and the oversized clear aperture defined and sent to RMI is shown in Figure 6b. The sides of the hexpyramid were measured on a Zygo® Interferometer to ensure it met the desired specifications. Masks in the shape of the clear aperture and circular masks larger than the beam footprint were used (Fig. 6c). In Figure 7 it is shown how the critical <λ/10 PV surface flatness specification was successfully met by each side of the pyramid.

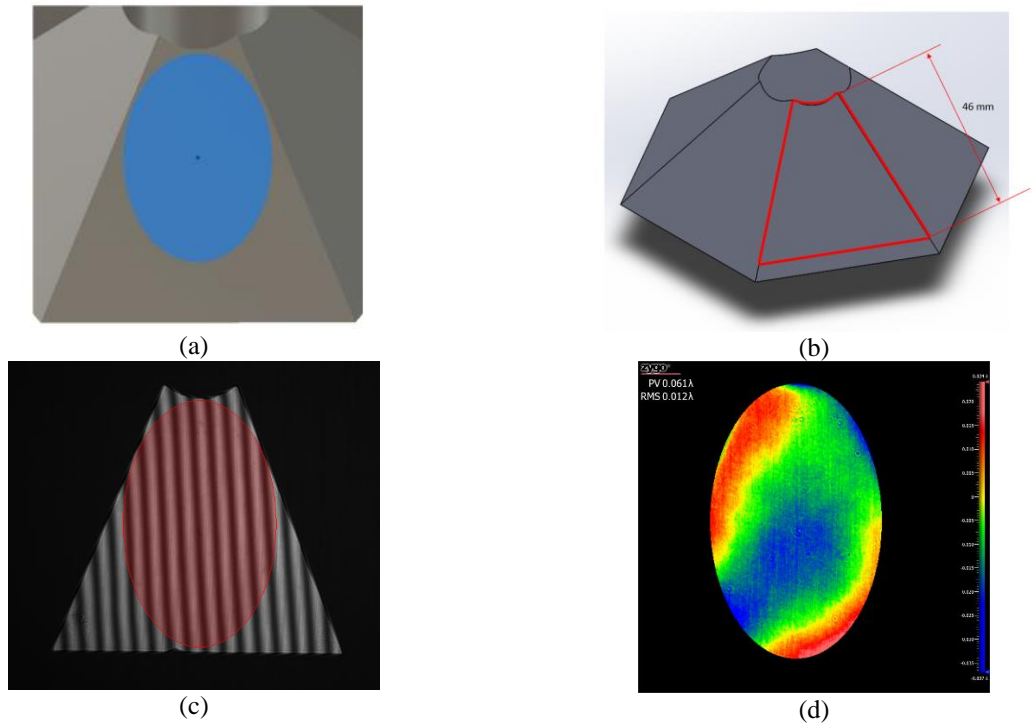

Figure 6. (a) Beam footprint (b) Clear aperture (c) Circular mask (d) Zygo® measurement

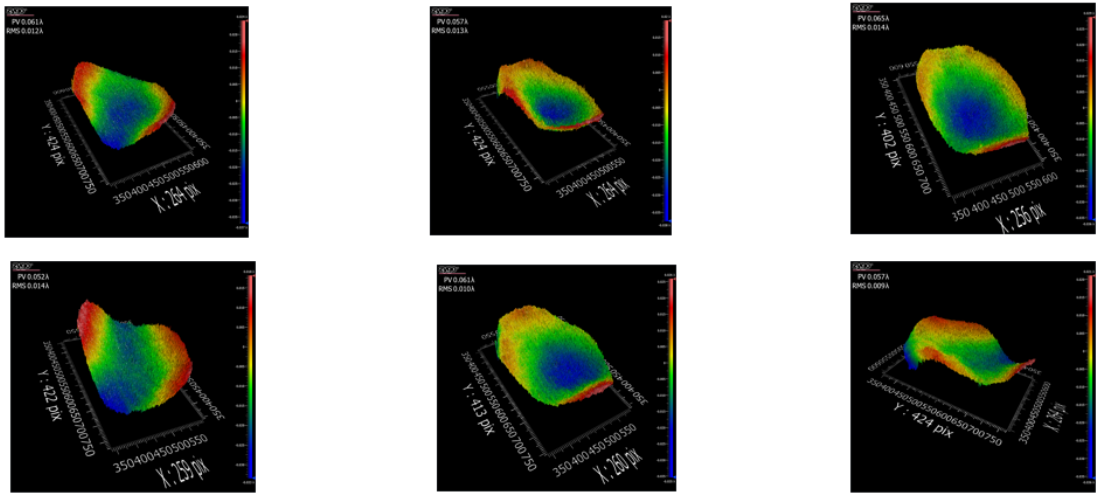

Figure 7. Interferometric measurements show that the PV spec of <λ/10 was met on each side of the hexpyramid

**2.2 Crossed Fold Mirrors and Splitting up the Pupil**

In order to split up the 7-segment GMT pupil, it will be passed onto the reflective hexpyramid. The six outer segments, or petals, will be sent outward in six different directions while the central segment passes through the center. Each segment will travel to its own flat fold mirror mounted on a piezoelectric piston/tip/tilt controller with +/- 60 μm of optical path difference stroke, then onto its own commercial 3,000 actuator Boston Micromachines (BMC) deformable mirror (DM) that will be employed for extreme wavefront control.

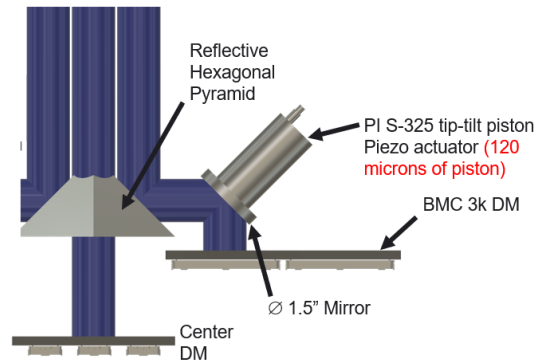

Figure 8. Hexpyramid sending individual pupil segment to other ExAO optics. Note: this design was rejected for the crossed fold mirror design shown in Figure 10

Originally, this design simply called for the fold mirrors, coated in protected silver, to be glued onto the piezoelectric controllers to fold the beam backwards, behind the hexpyramid. The complex refractive index of protected silver creates a phase shift that varies with wavelength between the s- and p-polarized components of the incident light (Fig. 9). So, when the beam segments are recombined, the resulting sum of the segment PSFs will be incoherent and low in Strehl. Thus, the design was modified for the mirrors to be crossed fold mirrors. Crossed fold mirrors are fold mirrors with a perpendicular plane of incidence to that of the prior reflection. A crossed fold mirror configuration proved to be critical in eliminating polarization aberrations[1]. Since the plane of incidence of a hexpyramid side is oriented 90 degrees to the plane of incidence of the corresponding fold mirror (a crossed fold), the s- and p-polarized light reflecting off the hexpyramid switch after reflecting off the fold mirror. This balances the net phase for a single incident angle and yields a linear variation of diattenuation. This allows the segment PSFs to combine coherently and produce a high-Strehl PSF (Fig. 10).

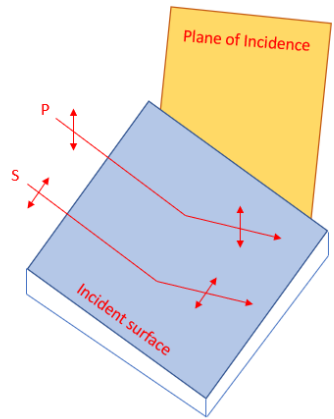

Figure 9. P-polarized light is parallel to the plane of incidence while s-polarized light is perpendicular to the plane of incidence. An incident unpolarized beam has less reflected p-polarized light. To maximize coherence, the next reflection must be off of a crossed fold mirror so that in essence, p- and s-polarized light switch, and the rest of the light matches in coherence. The beam remains unpolarized after the second reflection, as long as the coatings and angle of incidences are the same.

The Strehl ratio is defined as the ratio of the peak aberrated image intensity from a point source compared to the maximum attainable intensity. It is important for the protected silver coatings on the crossed fold mirrors to match that of hexpyramid to properly cancel the polarization aberrations and achieve a high Strehl ratio. It is relevant to note, the coating of the BMC 3k DMs does not need to match since they have an angle of incidence of 0.07°, thus no phase difference between s- and p-polarizations. A raytrace in Zemax OpticStudio® confirmed the cancelation of aberrations via Huygens PSF plots and Strehl plots (Figs. 10 and 11, Hedglen, et al. 2022)[1].

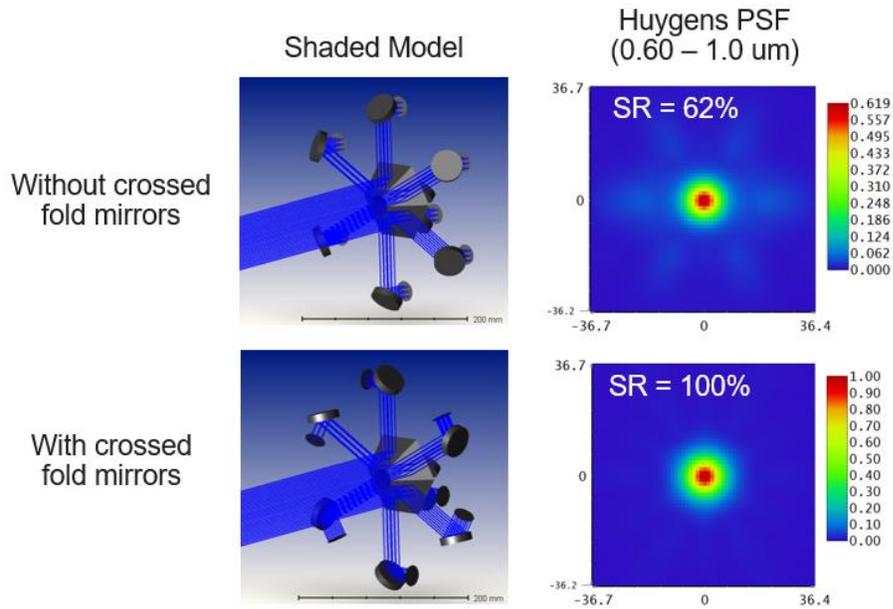

Figure 10. Comparison of polarization aberrations for cases of no crossed fold mirrors and crossed fold mirrors[1]

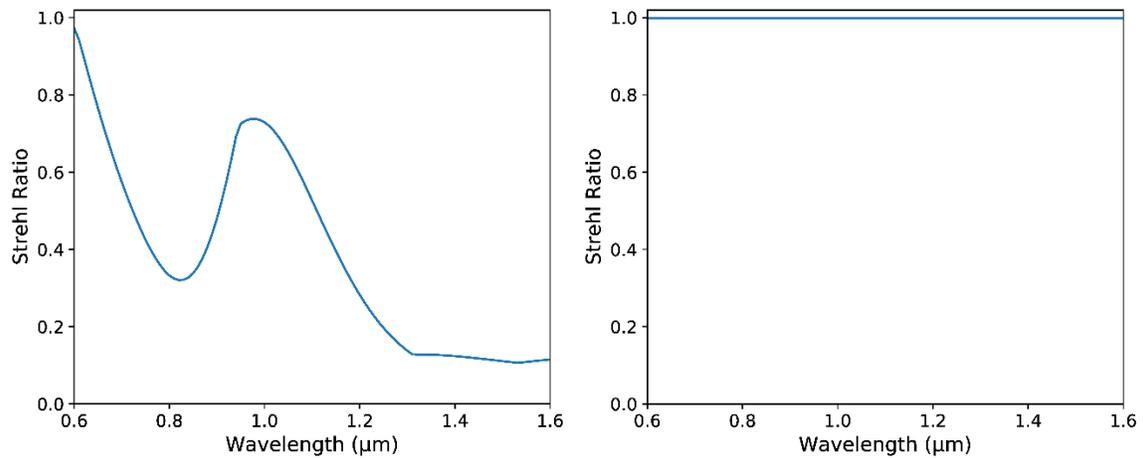

Figure 11. Strehl plots for non-crossed and crossed fold mirror configurations respectively[1]

It is important to note that the central segment must also pass onto crossed fold mirrors to achieve this high-Strehl PSF (Fig. 12).

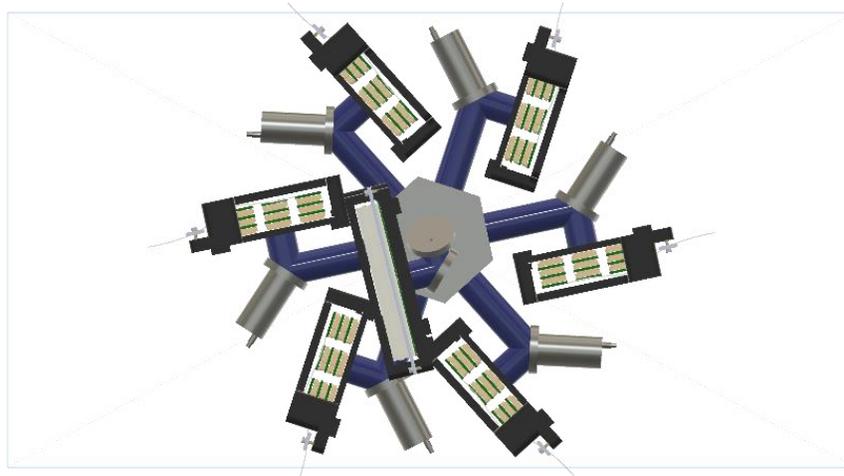

Figure 12. Back view of parallel DM

## 3. PARALLEL DM EXAO ARCHITECTURE

The expansive GMT pupil requires each segment to have a dedicated deformable mirror to reduce wavefront error to an acceptable level of < 50 nm RMS. Seven 3,000 actuator deformable mirrors working in parallel will be the equivalent of a 21,000 actuator DM[6]. This is what that could look like in a crossed fold mirror configuration:

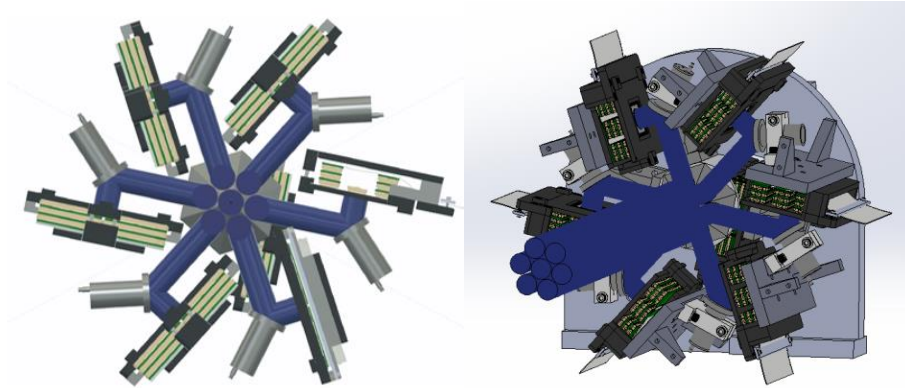

Figure 13. CAD renditions of parallel DM to be used on GMagAO-X

Before purchasing seven 3k DMs, at a very substantial cost, HCAT needs to prove that the GMT pupil can be split up, phased, and recombined coherently to create a high-Strehl PSF. This lead to the notion of "mock DMs." The mock DM has roughly the same dimensions of a real BMC 3k DM with an oversized flat square mirror for each segment (Fig. 14).

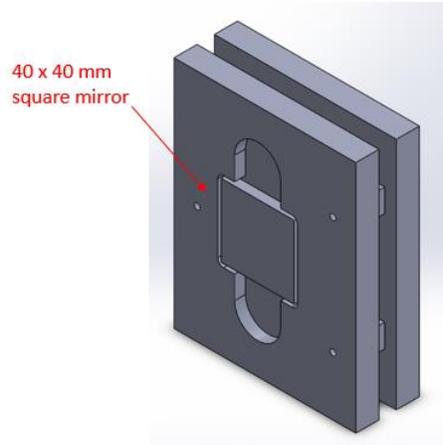

Figure 14. SolidWorks rendition of mock DM

Mock DMs will be put it place of the BMC 3k DMs to model the parallel DM system. The piezoelectric controllers will be used to add a 0.14 degree tilt to the beam so it can work in double-pass.

## 4. WATERWHEEL STRUCTURE

The next phase of this project was designing the opto-mechanical structure that will hold the hexpyramid, piezoelectric actuators, and mock-DMs that will be used to put the parallel DM idea to the test. This design was necessarily complex and compact, the entire structure is shaped a bit like a waterwheel to accommodate the crossed fold mirror design.

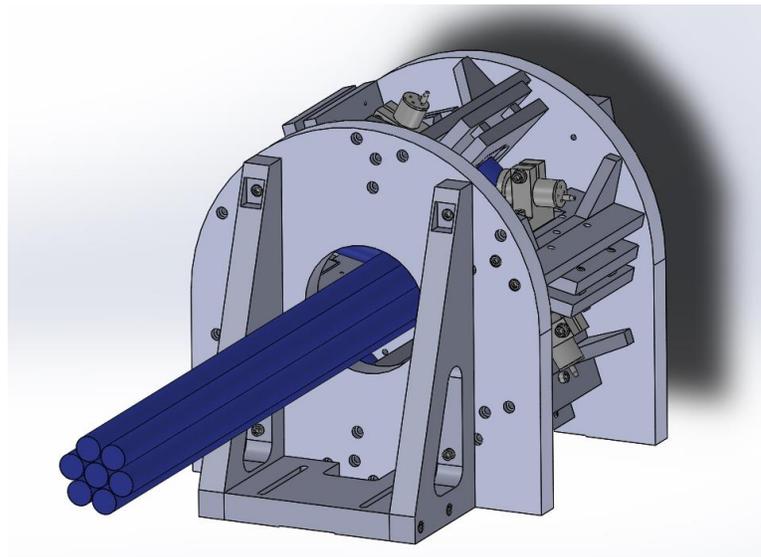

Figure 15. SolidWorks rendition of HCAT's parallel DM/waterwheel. The 3k DMs are replaced with mock DMs but otherwise this is identical to the GMagAO-X parallel DM/waterwheel.

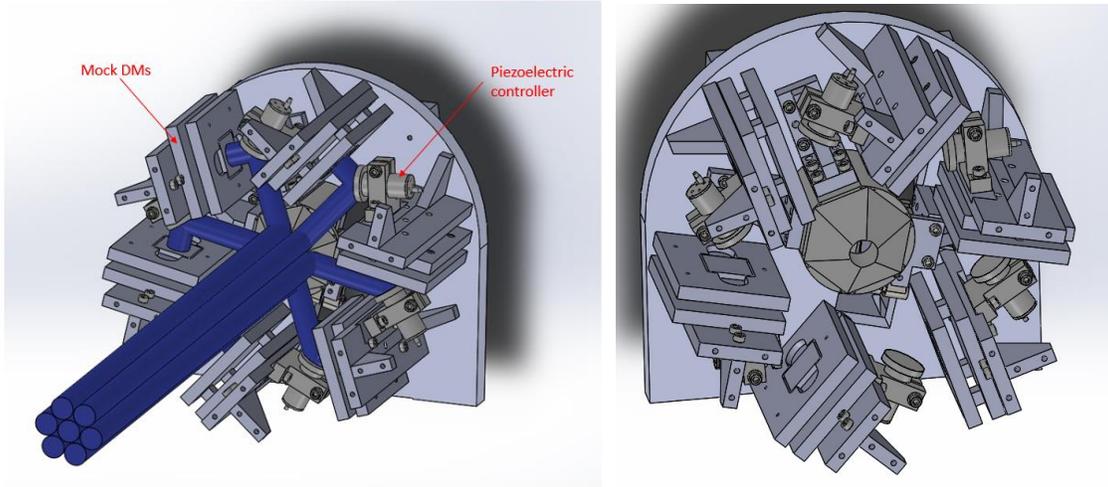

Figure 16. Open face views of Waterwheel structure

In order to achieve the high-Strehl PSF achieved during simulations, the crossed fold mirror configuration also needs to be implemented on the central segment. This is where the cylindrical mount, designed by senior graduate student Alex Hedglen, is utilized (Fig. 17).

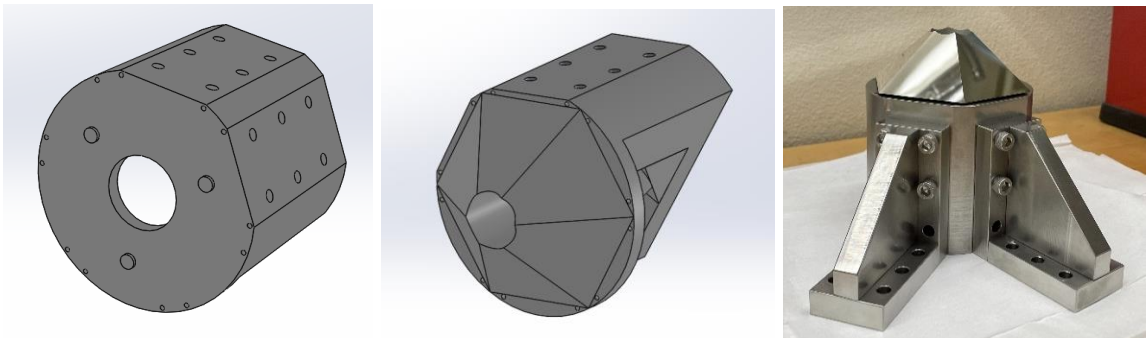

Figure 17. (left) CAD models of cylindrical mount, (right) fabricated hexpyramid mounted on cylindrical mount

The hexpyramid optic is glued onto the top of the cylindrical mount. Crossed fold mirrors are also bonded on the inside of the mount (Fig. 18).

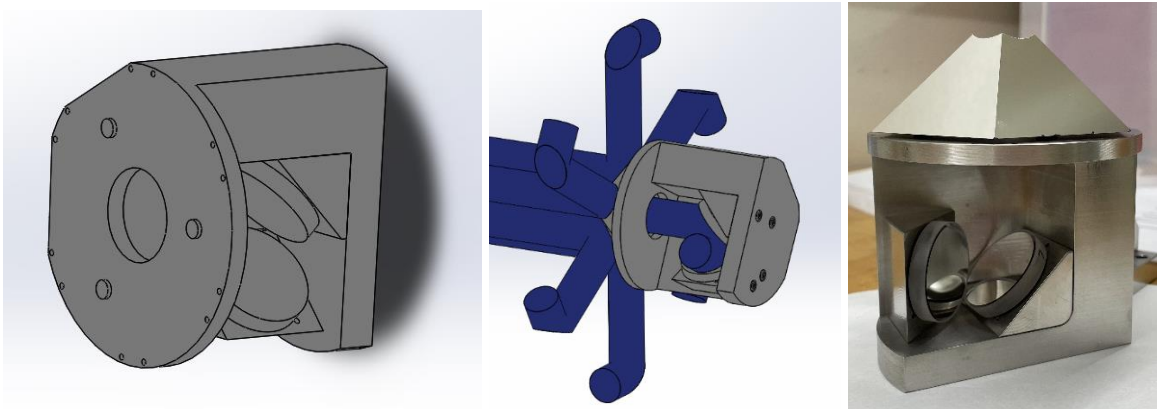

Figure 18. Crossed fold mirrors within cylindrical mount

These crossed fold mirrors allow light with zero net polarization errors to be sent to the central "DM." Due to space constraints, the central DM is not a mock DM with a piezoelectric controller but rather a 2" mirror in a vertical drive kinematic Polaris® tip/tilt mount (Fig. 19).

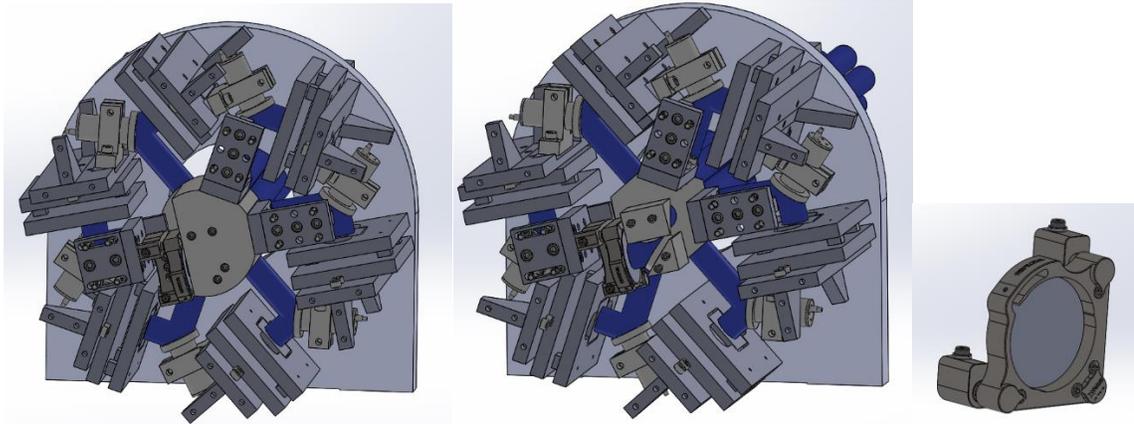

Figure 19. Back view of waterwheel and Polaris® kinematic mount

The tip/tilt drivers are accessible via a rectangular hole in the back of the waterwheel structure (Fig. 20).

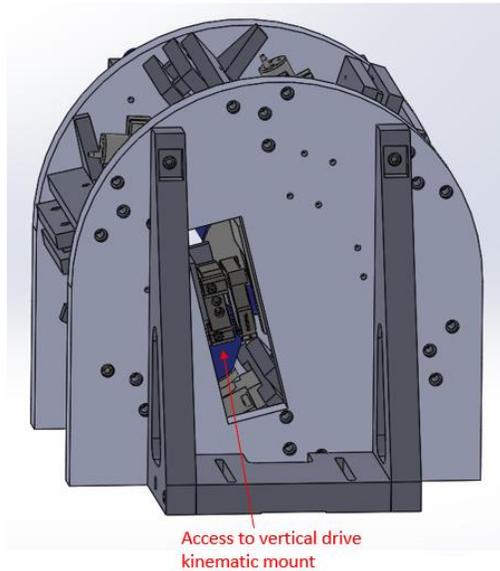

Figure 20. Rectangular hole on back of waterwheel structure

The waterwheel must be manufactured in invar to meet our thermal and vibration requirements. Invar has a particularly low coefficient thermal expansion (CTE, α), $1.2 \times 10^{-6}$ K$^{-1}$ (1.2 ppm/°C). This is compared to steel with a CTE of 11-15 ppm/°C and aluminum with a CTE of 20 ppm/°C. The total optical path length, $L_0$, from the hexpyramid to DM is 120 mm. With a temperature change of 1 degree Celsius, invar would produce a path length change of 0.288 μm. The factor of two comes from the fact that the system is in double pass.

$$\Delta L = \alpha L_0 \Delta T$$

$$\Delta L = (1.2 \cdot 10^{-6})(120 mm)(1°C)$$

$$\Delta L = 0.144 \ \mu m$$

$$\text{True } \Delta L = 0.288 \ \mu m$$

That same temperature delta with an aluminum structure would produce a total path length change of 4.8 μm. The difference may seem marginal in the lab but not at a telescope with ΔT's around 10°C where the OPD would be around 50 μm or 50-100λ. An invar parallel DM would only change in optical path length by about 3μm/night. Such slow changes can be tracked and completely removed by the PI S-325 controller since they have a +/- 60 μm stroke.

An FEA analysis performed by mechanical engineer, Fernando Coronado, showed the resonant frequency of the waterwheel structure is 785 Hz (Fig. 21). This is acceptable because the piezoelectric controllers will only operate at 100 Hz which is roughly 1/8$^{th}$ of the resonant frequency.

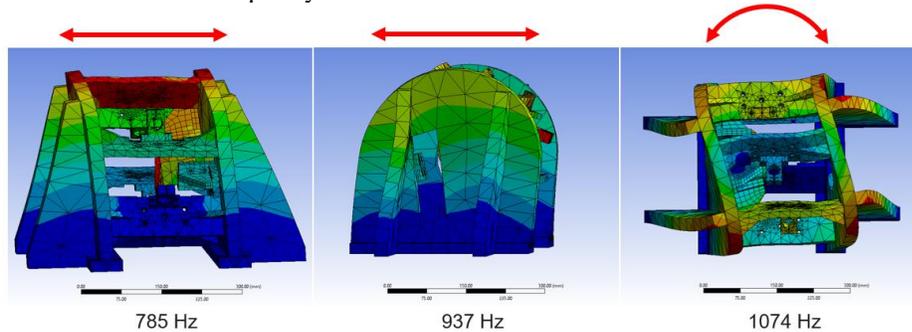

Figure 21. SolidWorks FEA Analysis. These are the lowest modes of the structure as built.

The waterwheel structure has recently been manufactured by the Machining & Welding Center at the University of Arizona (Fig. 22).

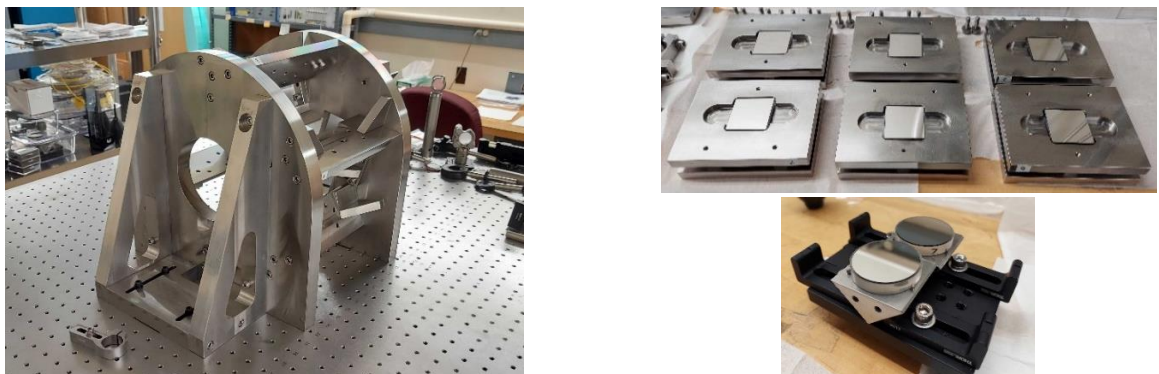

Figure 22. (left) Manufactured invar waterwheel structure placed on HCAT table, (top right) 40x40" flat square mirrors glued into mock DMs, (bottom right) crossed fold mirrors for central segment glued onto triangle mounts for cylindrical mount

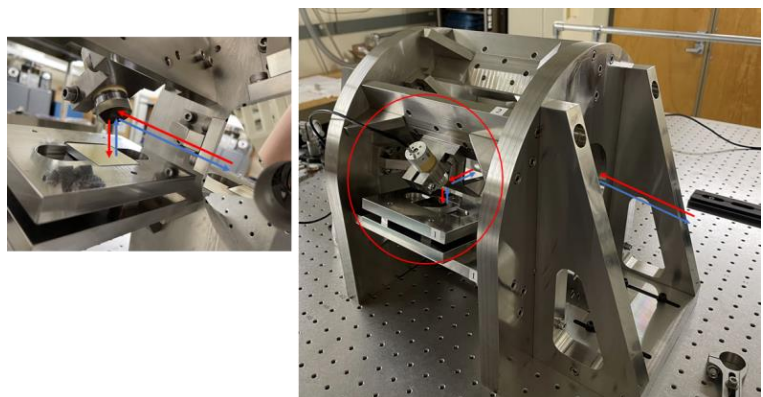

Figure 23. Test piezoelectric PI S-325 actuator and mock DM in place on waterwheel structure. This allows initial and on-going optical coherence tests of one outer segment with the central segment. The red ray traces the path of the light off the hexpyramid to the PI S-325 and to the mock DM. The blue ray traces the double pass path back to the hexpyramid.

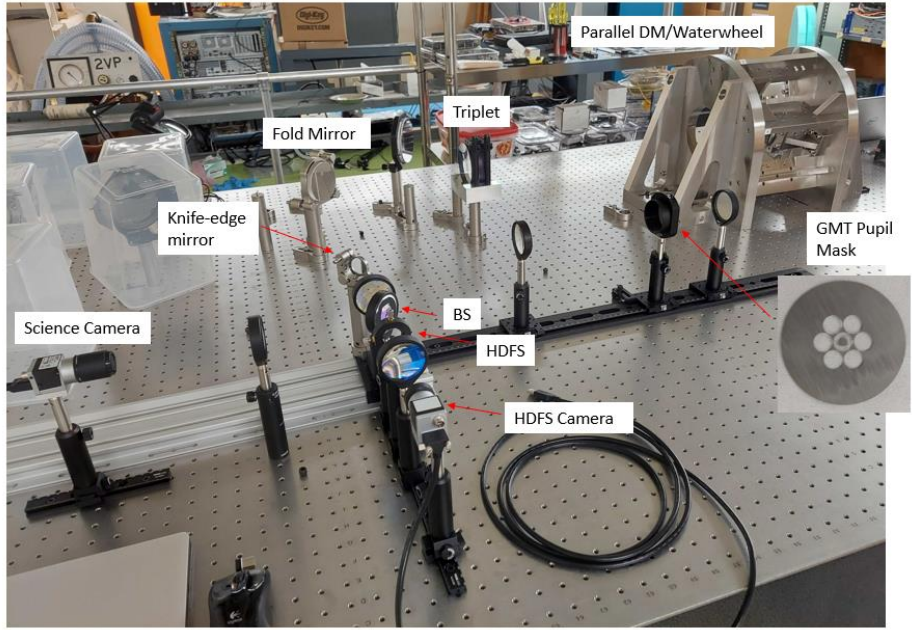

Figure 24. Parallel DM/waterwheel on HCAT table in an intermediate mode (optical design by Alex Hedglen)

The concept of the parallel DM will be tested by placing the waterwheel "parallel DM" structure onto the HCAT table (Fig. 25). The piezoelectric actuators will input a piston error onto one or more of the GMT segments. HCAT will feed this aberrated light into the existing MagAO-X instrument which will sense the segment piston using a holographic dispersed fringe sensor (HDFS) designed by Postdoc Sebastiaan Haffert[8,9]. This will create a closed loop system where the piezoelectric actuators can then be commanded to correct the injected piston error sensed by the HDFS. The 2k tweeter DM on MagAO-X can be used to inject turbulence into the system, as has already been demonstrated with p-HCAT[1]. When the mock DMs are eventually replaced with BMC 3k DMs, the pyramid wavefront sensor (PyWFS) on GMagAO-X will act as a slope sensor that will work in a closed loop with seven DMs to correct wavefront error[8].

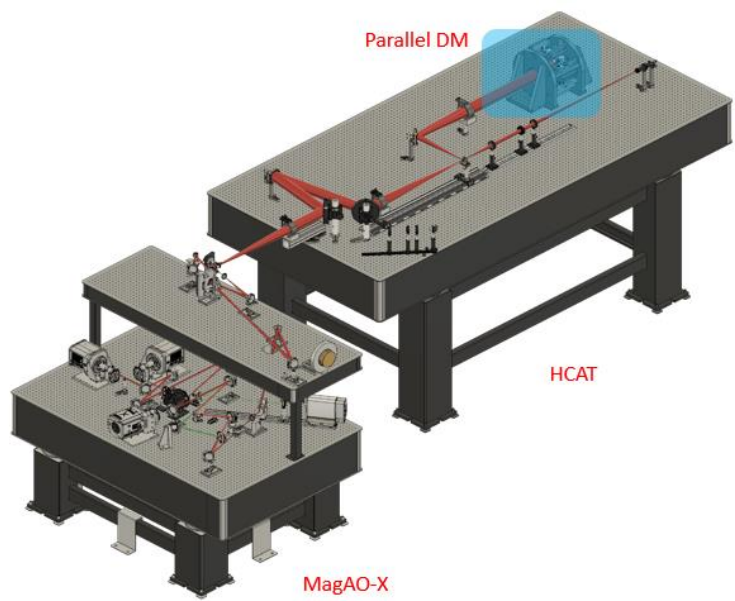

Figure 25. CAD rendition of HCAT feeding MagAO-X[1]

# 5. CONCLUSION

The entirety of this design was presented at the Final Design Review for the GMT Organization with world-class adaptive optics experts as reviewers. The design successfully passed FDR, giving the CAAO/HCAT team authorization to manufacture the structure and move forward with the upgraded testbed and thus manufacture the waterwheel. Now the invar waterwheel is complete and almost all of the optics have been glued in place. The only components yet to arrive are the six PI S-325 piezoelectric controllers that have been delayed to the early fall due to chip shortages. Once the controllers are in place, the next steps include validating the ability to slice the pupil and coherently recombine it in-phase and ultimately the entire parallel DM ExAO architecture. The entire HCAT will be used in closed loop with MagAO-X's PyWFS and HDFS with the GMT pupil. Thanks to the parallel DM, the six segments will be able to rapidly tip, tilt, and piston. The final phase of the HCAT testbed will be in summer 2023 when HCAT will be combined with MagAO-X to test the GMT's natural guide star (NGS) pyramid sensor (NGWS prototype)[10].

# ACKNOWLEDGMENTS

The HCAT testbed program is supported by an NSF/AURA/GMTO risk-reduction program contract to the University of Arizona (GMT-CON-04535, Task Order No. D3 High Contrast Testbed (HCAT), PI Laird Close). The authors acknowledge support from the NSF Cooperative Support award 2013059 under the AURA sub-award NE0651C. Support for this work was also provided by NASA through the NASA Hubble Fellowship grant #HST-HF2-51436.001-A awarded by the Space Telescope Science Institute, which is operated by the Association of Universities for Research in Astronomy Inc. (AURA), under NASA contract NAS5-26555. Maggie Kautz received an NSF Graduate Research Fellowship in 2019. Alex Hedglen received a University of Arizona Graduate and Professional Student Council Research and Project Grant in February 2020. Alex Hedglen and Laird Close were also partially supported by NASA eXoplanet Research Program (XRP) grants 566 80NSSC18K0441 and 80NSSC21K0397 and the Arizona TRIF/University of Arizona "student link" program. We are very grateful for support from the NSF MRI Award #1625441 (for MagAO-X) and funds for the GMagAO-X CoDR from the University of Arizona Space Institute (PI Jared Males) as well. This material is based upon work supported in part by the National Science Foundation as a subaward through Cooperative Agreement AST-1546092 and Cooperative Support Agreement AST-2013059 managed by AURA.